\documentclass[conference]{IEEEtran}
\IEEEoverridecommandlockouts
\usepackage{cite}
\usepackage{amsmath,amssymb,amsfonts}
\usepackage{algorithmic}
\usepackage{graphicx}
\usepackage{tabularx}
\usepackage{textcomp}
\usepackage{xcolor}
\usepackage{hyperref}
\usepackage{multirow}
\usepackage{subfigure}
\usepackage{float}

\def\BibTeX{{\rm B\kern-.05em{\sc i\kern-.025em b}\kern-.08em
    T\kern-.1667em\lower.7ex\hbox{E}\kern-.125emX}}
\begin{document}

\title{Automatic Generation of Digital Twins for Network Testing}

\author{
\IEEEauthorblockN{Shenjia Ding}
\IEEEauthorblockA{\textit{School of Computing Science} \\
\textit{University of Glasgow}\\
Glasgow, UK \\
2788155d@student.gla.ac.uk}
\and
\IEEEauthorblockN{David Flynn}
\IEEEauthorblockA{\textit{School of Engineering} \\
\textit{University of Glasgow}\\
Glasgow, UK \\
david.flynn@glasgow.ac.uk}
\and
\IEEEauthorblockN{Paul Harvey}
\IEEEauthorblockA{\textit{School of Computing Science} \\
\textit{University of Glasgow}\\
Glasgow, UK \\
paul.harvey@glasgow.ac.uk}
}

\maketitle

\begin{abstract}
The increased use of software in the operation and management of telecommunication networks has moved the industry one step closer to realizing autonomous network operation. One consequence of this shift is the significantly increased need for testing and validation before such software can be deployed. Complementing existing simulation or hardware-based approaches, digital twins present an environment to achieve this testing; however, they require significant time and human effort to configure and execute.

This paper explores the automatic generation of digital twins to provide efficient and accurate validation tools, aligned to the ITU-T autonomous network architecture's \textit{experimentation subsystem}. We present experimental results for an initial use case, demonstrating that the approach is feasible in automatically creating efficient digital twins with sufficient accuracy to be included as part of existing validation pipelines. 


\end{abstract}


\section{INTRODUCTION}

Autonomous networks represent the holy grail of network and service management, aiming to achieve self-configuring, self-optimizing, and self-healing capabilities with minimal human intervention~\cite{ITU-T2022ArchitectureNetworks}. Recent industry motivation for autonomous networks is driven by increasing complexity of networks, particularly with the advent of 5G and beyond, which demand more agile, scalable, and efficient management~\cite{SoftwarizationFor5G}. 

Traditional network management, which relies on human expertise and manual configurations, struggles to keep pace with the rapid advancements in network technologies and the diverse requirements of emerging services~\cite{enterpriseControl}. Recent technologies
have enabled software-based control of the network, paving the way for approaches such as Machine Learning (ML)~\cite{MLforNetwork} approaches to support dynamic and intelligent network operations. These technologies facilitate the network programmability, allowing real-time adaptation to changing conditions, resource optimization, proactive fault resolution, and enabling more granular control and optimization tailored to specific performance requirements.




\begin{figure}
    \centering
    \includegraphics[scale=0.5]{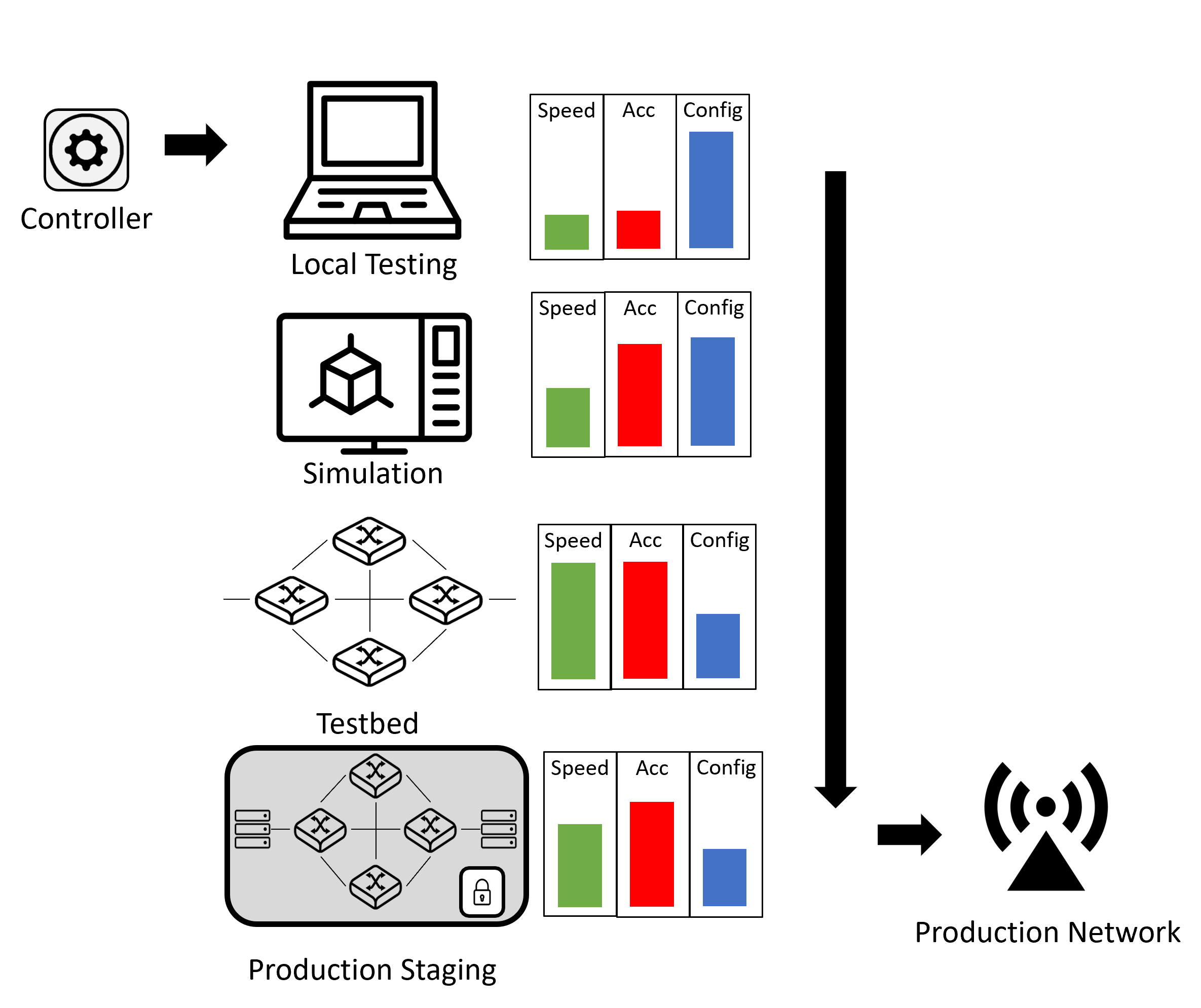}
    \caption{Telco Software Validation Tool Pipeline}
    \label{fig:telc-vt-pipeline}
\end{figure}

However, given that only 39\% of global software projects succeed~\cite{chaosreport}, holistic software deployment across the network exposes national critical infrastructure to significant levels of risk. Thus, telecommunication networks, like other safety-critical industries (e.g. aviation), require a significant validation/testing pipeline for network control software before it is considered for production deployment. This pipeline consists of multiple validation tools (VT), such as simulators, scripts, testbeds, or pre-production staging environments, where each VT has a trade-off between execution time (speed), accuracy, and configurability~\autoref{fig:telc-vt-pipeline}, leading to complex, inefficient, and often ineffective software validation (\autoref{sec:backgorund}). This excludes the time taken to prepare, configure, or schedule for each VT. This level of human involvement in the validation is a clear barrier to the telecoms industry's shift toward software-based operations and fully autonomous networks. 



Research on the potential of \textit{Digital Twin} (DT)\cite{NDT} technology has attracted widespread attention due to its ability to simulate end-to-end services through comprehensive resource management and planning\cite{DT6G}. This capability enables DT as a potential testing tool and enhances evaluation efficiency. 
However, based on current research, the development of DTs still faces the challenge that the generation process relies heavily on domain-specific expertise.

This paper presents an initial study on the automatic generation of digital twins for networks (\autoref{sec:AutoML}). Specifically, we explore the use of automated machine learning approaches to achieve data-driven, context-specific DT creation. To the best of our knowledge, this is the first work to do so. We demonstrate the approach through preliminary results using synthetic data, showing that the approach is efficient, sufficiently accurate, and viable (\autoref{sec:exp}).





\section{Background}
\label{sec:backgorund}

\subsection{Validation in Autonomous Networks}
\label{subsec:AN}


A key element of the ITU-T autonomous network architecture~\cite{itu-y3061} is the experimental evaluation subsystem, who's goal is to test and validate the software (\textit{controllers}) responsible for the network  management and operation. This concept of \textit{validation} is one of the key concepts in the approach. To ensure the effectiveness of the testing, environment modeling plays a crucial role in autonomous frameworks~\cite{CognitiveAN}.





\subsection{Current Validation Tools} 

Current approaches to validating network software are usually broken into a linear progression from software simulations to hardware testbeds and finally production deployments, each stage presenting its own challenges,~\autoref{fig:telc-vt-pipeline}.

\subsubsection{Hardware Tools}
Hardware testbeds consist of specialized purpose-built devices designed for network testing. However, these testbeds suffer from high costs and lack flexibility, as they cannot be easily reconfigured to accommodate new testing or monitoring functions~\cite{ProgrammableHardware}. As network requirements evolve, the difficulty of modifying or expanding testbeds limits scalability and slows down the implementation of updates.

\subsubsection{Simulation Tools}

Compared to hardware testbeds, simulation software offers greater flexibility and cost-efficiency. It eliminates the need for physical hardware, reducing both setup and maintenance costs. Unlike hardware platforms, constrained by physical limitations, simulation allows rapid parameter adjustments, enabling simulation of diverse network topologies and behaviors. In addition, simulations can execute multiple tests in a shorter time while ensuring consistent and reproducible conditions, making simulation a powerful tool for network experimentation. Common examples include OMNet++~\cite{omnet}, ns-3~\cite{ns3}, and Mininet~\cite{mininet}.

Although existing solutions meet the environmental modeling requirements for testing, they have limitations, particularly when faced with large numbers of parameters to be tested. Specifically, the execution time and human effort required in the simulator configuration act as a barrier to exploration.


\begin{table}[t]
    \centering
    \caption{Comparison of Network Testing Methods}
    \begin{tabular}[width=\linewidth]{lccc}
        \hline
        \textbf{Testing Method} & \textbf{Accuracy} & \textbf{Speed} & \textbf{Config Flexibility} \\
        \hline
        Physical Testbed & High & Low & Low \\
        Simulation  & Medium & Medium-High & High \\
        Digital Twin & Variable & High & Medium-High \\
        \hline
    \end{tabular}
    
    \label{tab:testing_comparison}
\end{table}

\textbf{Summary}:
As shown in \autoref{tab:testing_comparison}, different approaches involve trade-offs between accuracy, speed, and configuration flexibility. Physical testbeds offer high accuracy but suffer from low speed and limited flexibility, while simulations provide faster, more configurable testing with reduced accuracy.




\begin{figure*}
    \centering
    \includegraphics[width=0.8\textwidth]{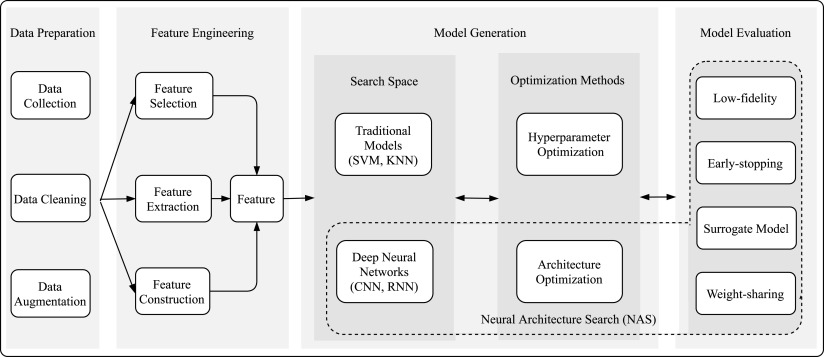}
    \caption{Framework of AutoML \cite{AutoML}}
    \label{fig:AutoML}
\end{figure*}

\subsection{Network Digital Twin}
\label{subsec:DT}

A Digital Twin (DT) is digital representation of a process, asset, or physical object~\cite{DTparadigm
}. 
Unlike traditional network testing described above, DT technology offers a transformative approach, integrating diverse network behavior data, constructing accurate models, and establishing mappings between real-world objects and their virtual counterparts~\cite{NDT}. This enables advanced prediction, emulation, analysis, and dynamic simulation of network, significantly enhancing efficiency and scalability in network testing. DT technology provides a solution to overcome existing methods' limitations by better reflecting network behaviors and pave the way for more responsive and efficient network testing solutions.

In the context of networks, DT is already enabling the development of more efficient network control and management tools for modern communication networks \cite{NDT}. Specifically, they are 
leveraging data from the network itself, enabling extensive testing without disrupting live systems\cite{DTenabledExpTest}\cite{DTbasedTesting}\cite{DTasTesting}. The ITU-T has also highlighted several use cases where DTs serve as critical platforms for experimentation, ensuring safe, scalable, and efficient validation~\cite{ituUsecase}.


In this work, we use DT to describe a data-driven model of a network, corresponding to a \textit{functional} DT~\cite{DTtoImproveQoS}, see \autoref{sec:AutoML}.

\subsection{Machine Learning-based DTs}

Machine Learning (ML) is an increasingly common approach for creating DTs in network-related applications. 
One example is in enhancing simulation accuracy and efficiency in network environments, with techniques like deep learning and reinforcement learning (RL) enabling high-speed, high-accuracy simulators for tasks such as network routing optimization \cite{DataDrivenRouting}. ML is also widely applied to improving Quality of Service (QoS) prediction and overall network performance, with models like Graph Neural Networks (GNNs), Spiking Neural Networks (SNNs), and ensemble models optimizing delay prediction \cite{DTtoImproveQoS} and CNN-GRU architectures being used for network capacity prediction \cite{PowerInfoNet}. 
Furthermore, ML-driven DTs support intelligent decision-making by facilitating resource management, network optimization, and real-time monitoring, as demonstrated in Deep Reinforcement Learning (DRL)-based synchronization and prediction models \cite{DTManager}. 
Finally, ML is instrumental in large-scale data processing, enabling real-time big data management and network life cycle monitoring \cite{NetPlanDesign}, ensuring that DTs can effectively handle vast amounts of network data for dynamic and evolving systems.



While ML models often focus on direct decision-making, DTs are generally designed to serve as neutral physical and functional mapping. Despite this distinction, their role under some specific cases\cite{DataDrivenRouting}\cite{PowerInfoNet}\cite{DTManager} demonstrates a potential capability that can be used to support the \textit{generation} of DTs. Recent studies have shown that ML can effectively support the generation of DTs by modeling complex, dynamic behaviors~\cite{DTfor5G, DT6G}. As such, ML-enabled DTs are emerging as powerful tools for anomaly detection, adaptive testing, and performance optimization in autonomous networks.



     \subsection{Problem}


Despite its potential, the broader application of ML in DT generation faces several barriers. Model development often requiring expertise that many network engineers lack limited and insufficient practical guidance~\cite{shouldNetAI}. The rapid evolution of ML also demands continuous learning, making tests highly knowledge-intensive and resource-demanding.

Key challenges include: data quality, where noise, missing values, and anomalies can significantly degrade prediction accuracy; feature engineering, where poor feature selection can lead to overfitting, bias, and poor generalization; and model selection and optimization remain complex, requiring extensive experimentation and computational resources, with advanced models often introducing further risks of errors.

\section{AutoML-enabled Automated Generation}
\label{sec:AutoML}

To address the challenges of creating DTs, this work explores the automated generation of DTs using Automated Machine Learning (AutoML). The goal is to be capable of handling various testing scenarios and processing the associated data with minimal manual effort. AutoML automates key steps, such as data preprocessing, feature engineering, model selection, and hyperparameter tuning (\autoref{fig:AutoML}) significantly reducing development time and human expert involvement.




By simplifying the training process, AutoML makes ML more accessible to ordinary developers and domain experts. Traditional model development is often labor-intensive, involving manual tuning and iterative testing, which hinders scalability. AutoML streamlines this process by efficiently evaluating multiple models and configurations, accelerating model generation while maintaining high performance.

As AutoML reduces training costs, lowers technical barriers for users, and automates comparison of multiple models' performance, we believe that AutoML may help automate the process of generating DTs for validation.

Unlike current \textit{system-level approaches}~\cite{DTmodelingOverview} to DTs, which aim for holistic representations, we instead focus on what we term \textit{Unit Twins}, with the aim of capturing specific network functionalities or components in a particular context to enable testing and validation. However, a network testing environment often requires evaluating multiple functionalities simultaneously, meaning that several Unit Twins may be needed to serve within the same network. We believe that such unit twins can also enable system-level testing and validation, however, that is beyond the scope of this work and left for future consideration. Like any other VTs (\autoref{fig:telc-vt-pipeline}), we consider that DTs would be integrated into the current testing and validation pipeline, as opposed to a perfect, stand-alone tool.



Given the relatively simple, array-based dataset used in this study (\autoref{sec:exp}), we focus on two AutoML tools: \textit{AutoGluon} and \textit{Auto-sklearn}, both well-suited for tabular data.

\textbf{AutoGluon}~\cite{Erickson2020AutoGluon-Tabular:Data} is an open-source framework that automates the ML pipeline, supporting tasks such as classification and regression. It offers a user-friendly interface and performs well on large-scale datasets.

\textbf{Auto-sklearn}~\cite{Feurer2019Auto-sklearn:Learning} builds on scikit-learn and focuses on traditional ML, using Bayesian optimization and meta-learning for model selection and hyperparameter tuning.

These frameworks differ in scalability, model selection strategies, and computational efficiency. We evaluate both to examine how AutoML choices influence the performance and effectiveness of DT generation.

\section{Preliminary Experiments}
\label{sec:exp}


The objective of this initial experimental study is to evaluate the feasibility of using AutoML to automatically generate a DT of the network from data alone. Specifically, can AutoML simplify and accelerate the creation of accurate and reliable digital replicas of network environments, and produce efficient and accurate DTs of the network.

\begin{figure}
    \centering
    \includegraphics[width=\linewidth]{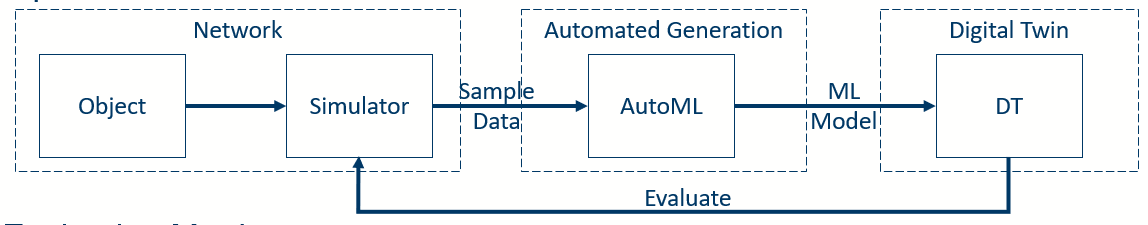}
    \caption{Experiment Pipeline}
    \label{fig:exp-pipeline}
\end{figure}


\subsection{Network Scenario}
\begin{figure}
    \centering
    \includegraphics[width=0.6\linewidth]{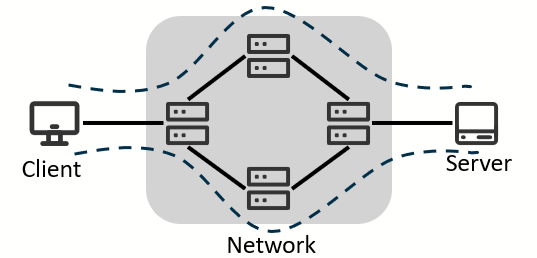}
    \caption{Network Topology}
    \label{fig:topo}
\end{figure}

We consider a simple network topology, \autoref{fig:topo}. The network is a diamond topology consisting of two paths, where the link bandwidths and queue size in the routers are configurable. Each links bandwidth ranged from 25Mbps to 125Mbps, and each router queue size ranged from 50 to 150. This simple topology allows evaluation of how different traffic conditions impact network latency. Upon the network, a simple application was used. Specifically, a 100MB file was requested by the client to the server and then downloaded through two different network paths while adjusting configuration parameters such as queue size and bandwidth. The blue dotted line in \autoref{fig:topo} shows each network path.

The objective for the generated DT is to capture the relationship between network parameters and the resulting latency on each path.
Such a DT would enable testing and validation of a traffic engineering network controller.

 \subsection{Experiment Setup}


\subsubsection{Data Generation}

Mininet~\cite{mininet}, a network emulator, was used to create the above scenario and generate a dataset. A script was used to automate the configuration and data generation of all possible combinations of link bandwidths and router queue sizes, as well as resulting path latency times. 


From this dataset, training and testing subsets were sampled. In alignment with the ultimate goal of reducing the time cost for network testing, we did not utilize the entire dataset for training. Instead, we used an interval sampling strategy to ensure that key characteristics of the network behavior were preserved. This approach was intended to balance between data volume and the retention of essential patterns, thereby enhancing the generalizability of the automated generation.

\subsubsection{AutoML Training}

We utilized the sampled data to train models using AutoGluon and Auto-sklearn. For  AutoGluon, we configured the task as a regression problem, aligning with our objective of predicting network latency. To ensure efficiency, we imposed a time limit of 300s 
on the entire training process and saved all intermediate models generated. After experimenting with several preset configurations in AutoGluon, we selected the 'good\_quality' setting, which offers a balance between predictive accuracy and computational efficiency. This setting offered fast inference speeds and lower disk usage compared to the 'high\_quality' option, making it suitable for applications where rapid generation are critical.

For the Auto-sklearn experiments, we configured the task as a regression problem and retained the default settings for all other parameters, allowing us to evaluate the out-of-the-box performance of Auto-sklearn without extensive customization, providing a baseline for comparison with AutoGluon.



\subsection{Result}


To evaluate the feasibility of automatically generating DTs via AutoML we consider accuracy of generated DT, time to generate a DT, DT execution time.




\subsubsection{Accuracy}

We trained multiple models using AutoGluon, leveraging its automated model selection process to identify the most suitable model. The training results, including test scores , are shown in \autoref{tab:AutoGluonResult}. 

Here we can see
the performance of different models on datasets from the two distinct paths. Notably, due to differences in the patterns of configuration variations during data generation, the dataset from Path 1 exhibits more regularity after sampling. 
In contrast, the dataset from Path 2 demonstrates a relatively more random distribution of these parameters. This distinction in data characteristics between the two paths provides valuable insights into how the inherent structure and variability of the datasets influence model performance.


\begin{figure*}[h]
    \centering
    \subfigure[Raw latency Dataset]{
        \includegraphics[width=0.3\textwidth]{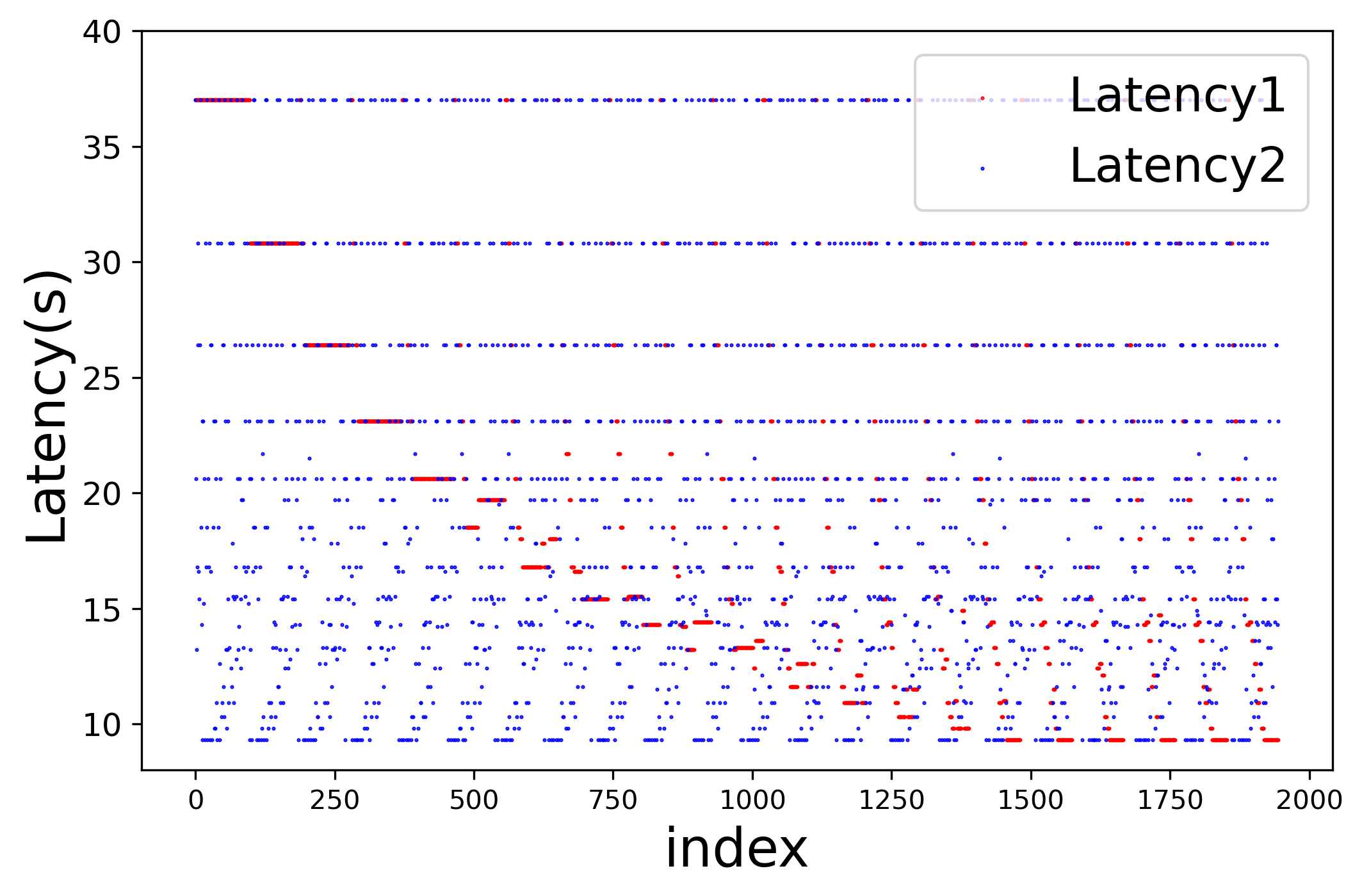}
        \label{fig:latency-raw-1}
    }
    \subfigure[Noised latency Dataset]{
        \includegraphics[width=0.3\textwidth]{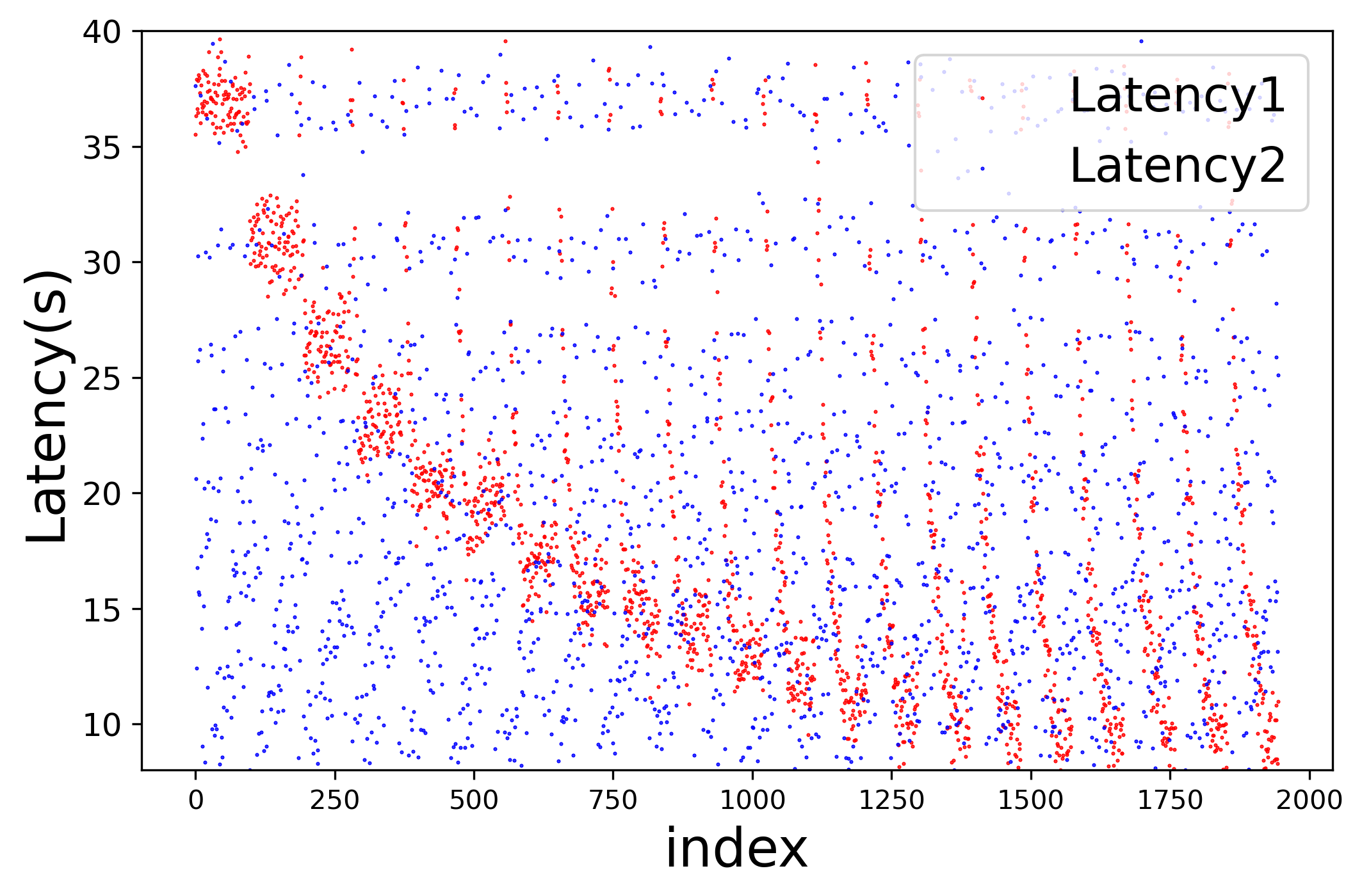}
        \label{fig:raw-noise}
    }
    \subfigure[Cleaned latency Dataset]{
        \includegraphics[width=0.3\textwidth]{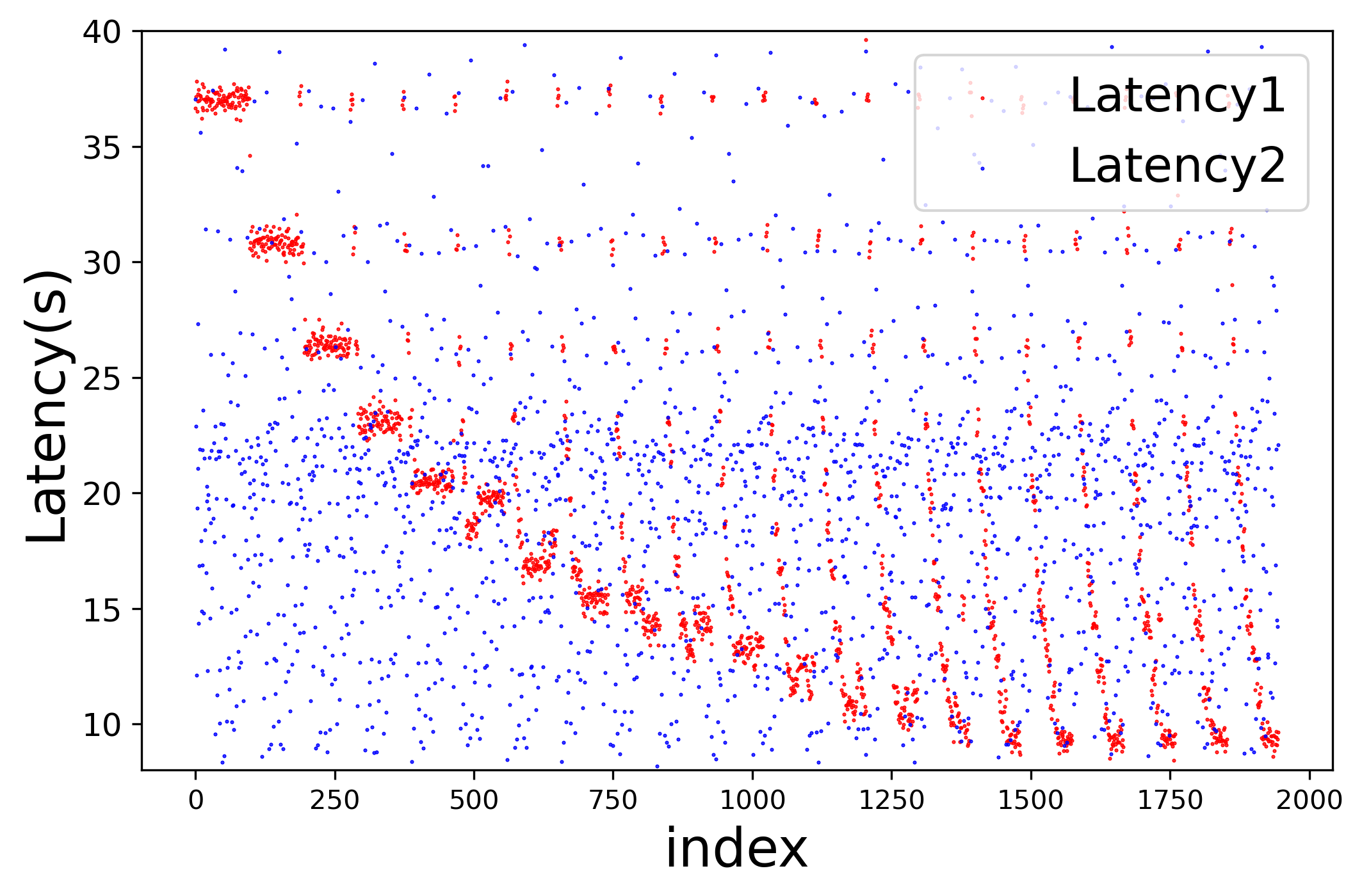}
        \label{fig:raw-savitzky-golay}
    }
    \subfigure[Accuracy (Original)]{
        \includegraphics[width=0.3\textwidth]{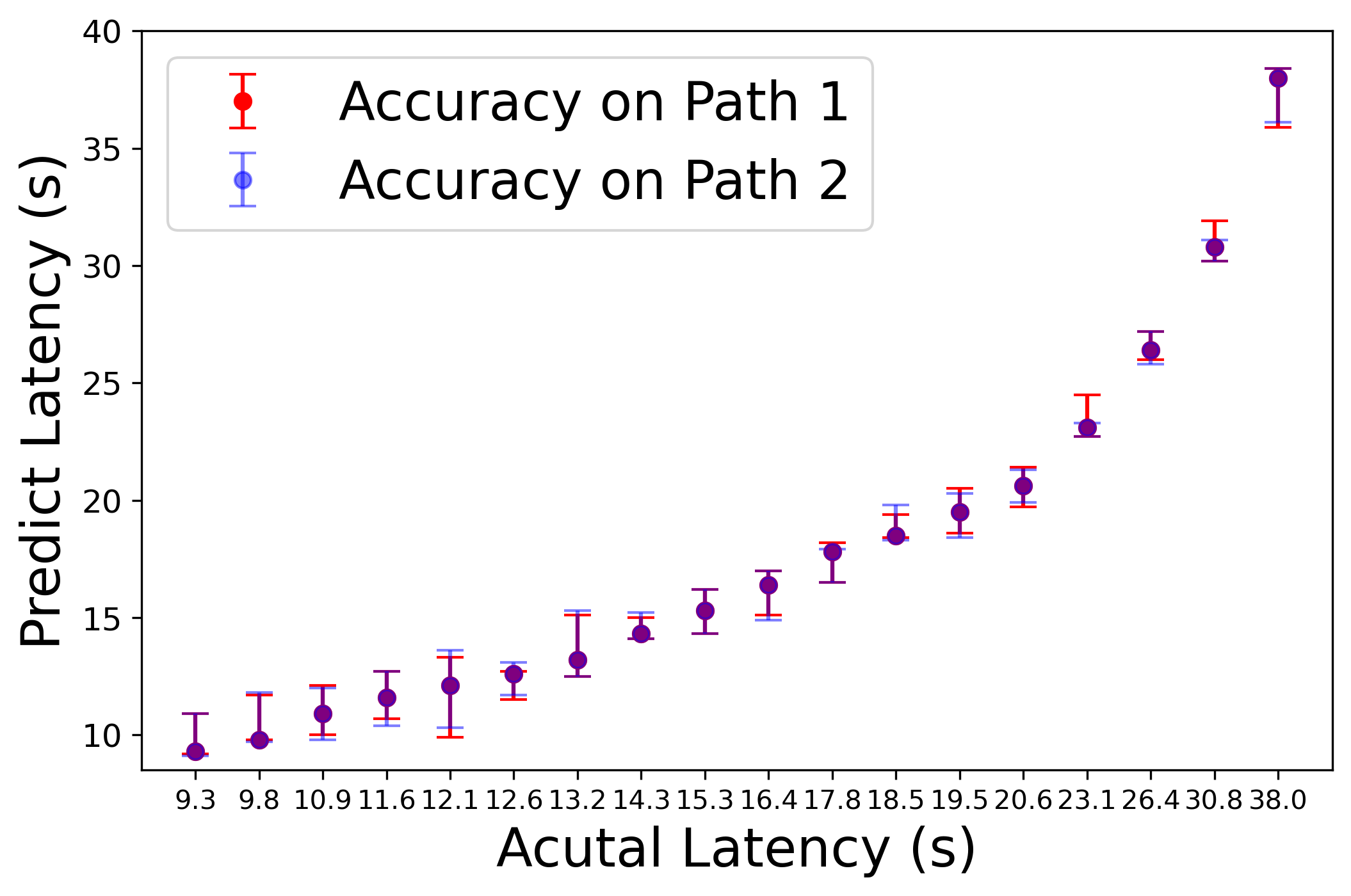}
        \label{fig:latency-raw-2}
    }
    \subfigure[Accuracy (Noised)]{
        \includegraphics[width=0.3\textwidth]{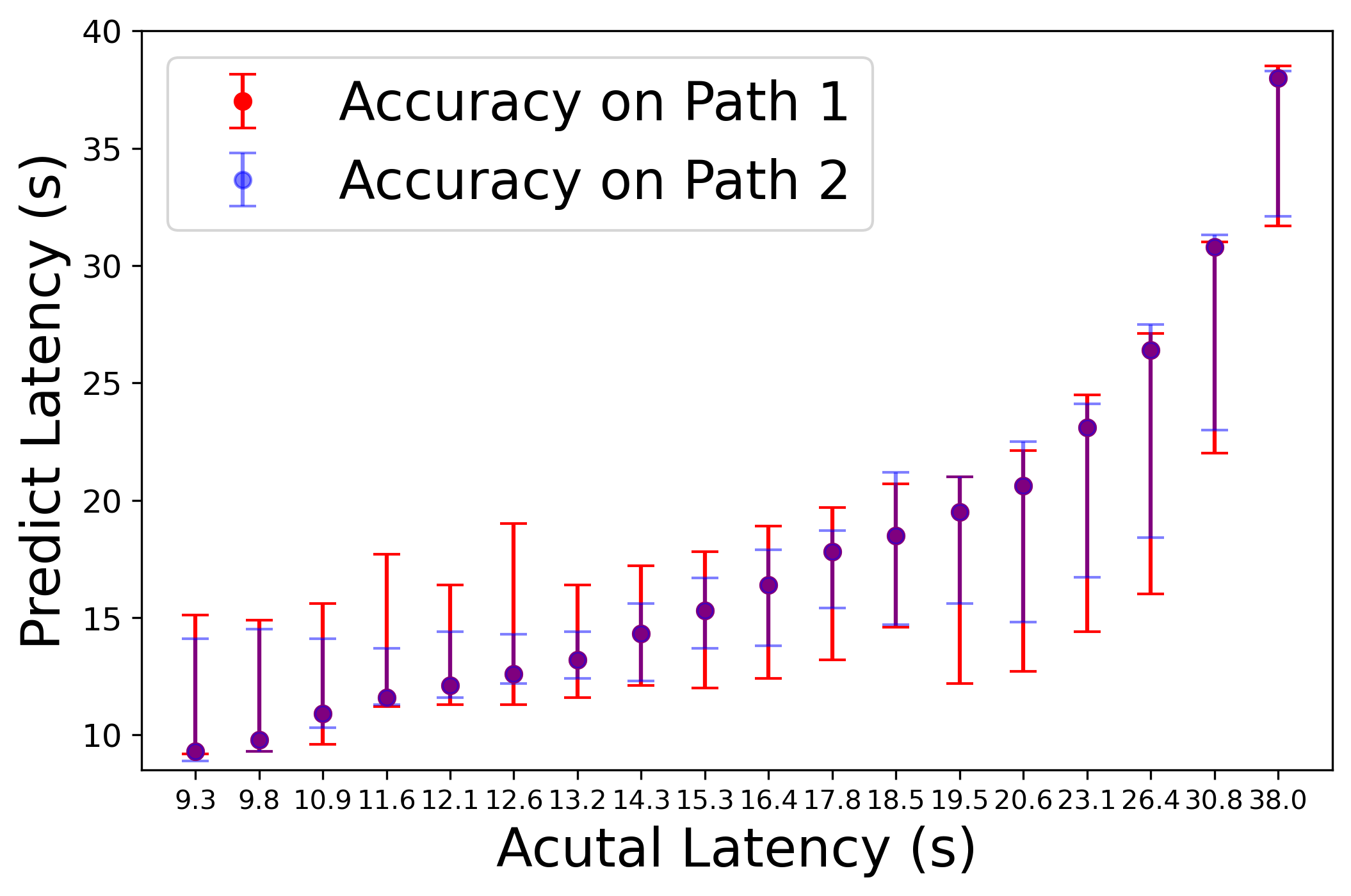}
        \label{fig:latency-noise}
    }
    \subfigure[Accuracy (Cleaned)]{
        \includegraphics[width=0.3\textwidth]{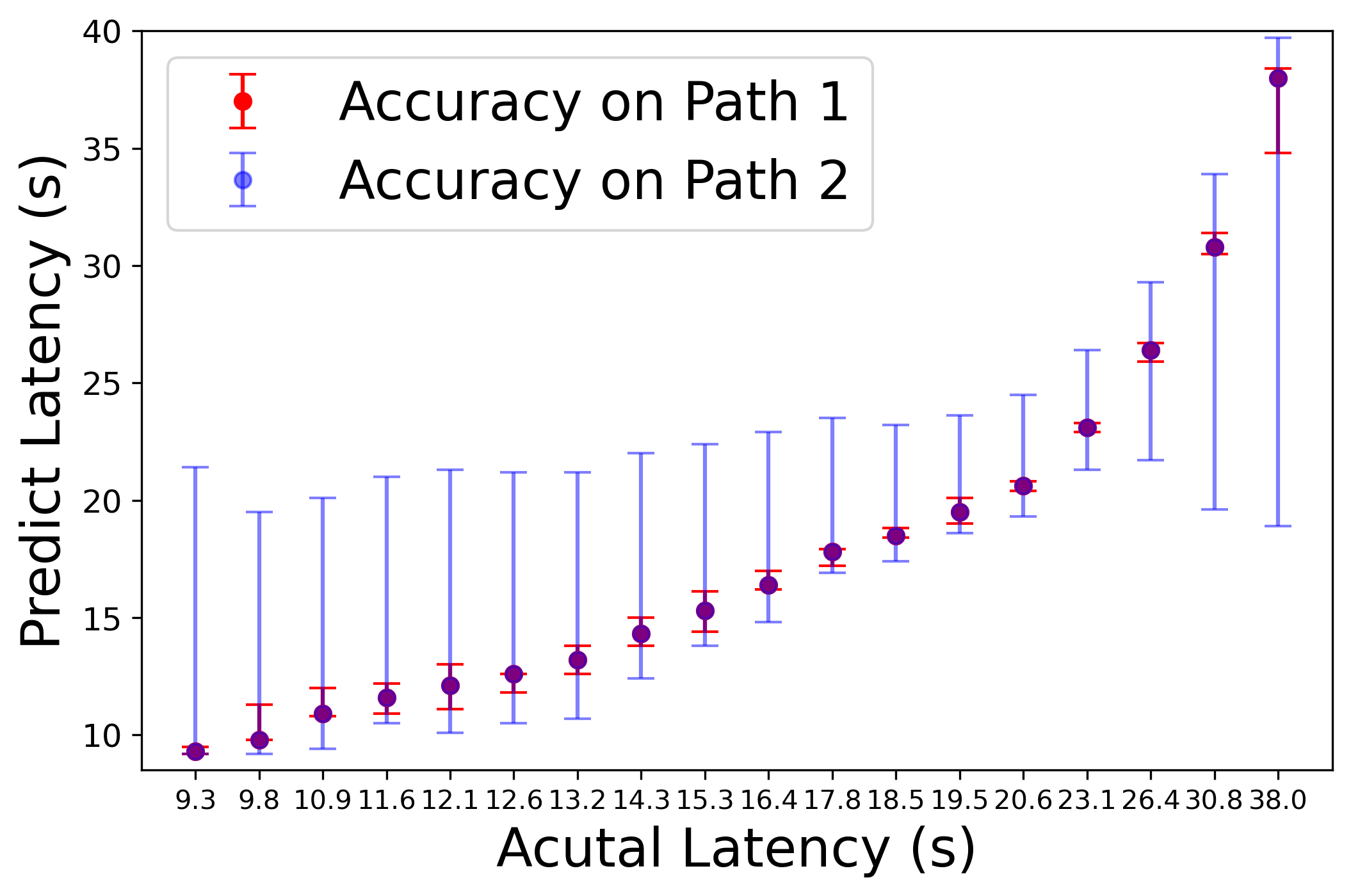}
        \label{fig:latency-clean}
    }
    \caption{Comparison of dataset processing and its impact on model accuracy. Blue values are for Path 1, red values are for Path 2.}
    \label{fig:data processing}
\end{figure*}

\begin{table}[]
    \caption{Result of Each Candidate Model in AutoGluon}
    \centering
    \begin{tabular}{l|c|c|c|c}
         \multirow{2}{*}{\textbf{Model}} & \multicolumn{2}{c}{\textbf{Path 1}} & \multicolumn{2}{c}{\textbf{Path 2}} \\
         \cline{2-5}
         & \textbf{Accuracy} & \textbf{Time(s)} & \textbf{Accuracy} & \textbf{Time(s)} \\
         \hline
         LightGBM & 99.63\% & 0.42 & 84.43\% & 0.41 \\ \hline
         ExtraTreeMSE & 99.71\% & 1.94 & 85.02\% & 2.05 \\ \hline
         RandomForestMSE & 99.75\% & 2.62 & 84.91\% & 2.33 \\ \hline
         CatBoost & 99.78\% & 23.53 & 84.46\% & 0.71 \\ \hline
         WeightedEnsemble\_L2 & 99.79\% & 25.44 & 91.57\% & 1.48 \\ \hline
         XGBoost & 99.78\% & 0.75 & 91.57\% & 0.52 \\ \hline
         Total AutoML Time (s) & \multicolumn{2}{c}{56.7} & \multicolumn{2}{c}{7.5} \\ \hline
    \end{tabular}
    
    \label{tab:AutoGluonResult}
\end{table}

Using the same training and validation datasets, we conducted experiments with both AutoGluon and auto-sklearn, and the results are presented in the \autoref{tab:GenTime}. The high accuracy achieved by both tools indicates that they are capable of generating DT effectively for the simple network topology test scenarios in our current experiments. However, determining which tool is more suitable for practical applications will require further investigation. In future work, we plan to evaluate their performance under more complex testing requirements and scenarios to provide a comprehensive comparison. 

\begin{table}[h]
    \centering
    \caption{Best Results of Different AutoML tools}
    \begin{tabular}{c|c|c}
        \hline
        Model & Accuracy on Path 1  & Path 2 \\ 
        \hline
        AutoGluon & 0.99 & 0.94 \\ 
        Auto-sklearn & 0.99 & 0.92 \\ 
        \hline
    \end{tabular}
    \label{tab:GenTime}
\end{table}

Using either AutoML framework, the results indicate that a DT \textbf{can} be automatically created from data which captures the relationship between network configuration parameters (queue size and bandwidth) and latency with high accuracy. 

\subsubsection{Execution Time}
Execution times for Mininet and the generated DT were compared for the same network configuration based on the above scenario. The DT execution time (0.0167s) was over \textbf{500x faster} than Mininet (8.5316s). This \textit{significantly} lower execution is a clear advantage of the DT and presents an opportunity for significant scaling in network testing to meet the demands of autonomous networks.

Furthermore, when considering the total time required—including both the training data collection via simulation and the model training process—the time-effectiveness of the DT-based approach becomes even more evident. To evaluate a large-scale dataset consisting of 194,481 samples, the DT was trained using only a carefully designed subset of 400 samples, achieving a prediction accuracy of approximately 98 \%. The complete DT pipeline, including simulation-based data collection, model training and DT testing, required approximately 3.4 hours. In contrast, conducting full-scale testing using Mininet for all samples would require over 900 hours of continuous simulation time. This dramatic reduction in total execution time—by a factor of over 260×—highlights the superior efficiency of the proposed method. By significantly lowering the computational and temporal costs associated with large-scale testing, ML-enabled DTs offer a highly practical and scalable solution for evaluations in autonomous network environments.



    


\subsection{Data Quality} 

Given the simple scenario and perfect data quality generated from the simulators, we wanted to understand how well the AutoML frameworks could handle noisy data, as would be generated from real networks, and the resulting effect on the generated DT.


\subsubsection{Experimental Setup}

Working with the same scenario as before (\autoref{fig:topo}), we now generate a dataset using ns-3; this was due to the time overhead of generating data with Mininet. The raw latency values are seen in Figure~\autoref{fig:latency-raw-1} and provide a baseline of comparison.
Next, to simulate real-world imperfections, we added  Gaussian noise~\cite{gaussiannoise} with a standard deviation of 1 and a mean deviation of 0 to the raw latencies, Figure~\autoref{fig:raw-noise}. Finally, we applied  a Savitzky-Golay filter~\cite{savitzky} to the noised dataset to account for a scenario where a known de-noising approach can be pre-applied before AutoML, Figure~\autoref{fig:raw-savitzky-golay}.

\subsubsection{Result}

Figure~\autoref{fig:latency-raw-1}, ~\autoref{fig:latency-noise}, ~\autoref{fig:latency-clean} show the per-path latency predictions for the raw, noisy, and de-noised data respectfully. Red values are for path 1 and blue values for path 2. The 
AutoGluon framework is used. 




Similar to \autoref{tab:AutoGluonResult}, results vary by path.
Specifically, Path 1 shows a more consistent and overall decline in performance across the dataset, which can be attributed to the sequential configuration changes made during the generation of simulation data. On the other hand, Path 2 should display periodic performance fluctuations in the complete dataset 
but the sampling interval and Path 2's periodicity misaligned, making feature less noticeable.


The current denoising approach does not universally enhance model performance. In fact, Path 2 performance on the cleaned dataset is even worse than its performance on the noised dataset, suggesting that the filtering process may have removed some meaningful patterns or introduced artifacts that negatively impacted the model. However, the Path 1 performance shows an improvement, with the DT outperforming its counterpart trained on the original dataset, highlighting the nuanced impact of data pre-processing on model performance.


These findings underscore the importance of tailoring data pre-processing techniques to the specific characteristics of the dataset and the problem for network testing requirements. While noise reduction by filtering can be beneficial in certain contexts, it is not a one-size-fits-all solution. Future research should focus on developing more adaptive pre-processing methods to better support AutoML tools in handling diverse and imperfect datasets.






\section{Future work}
Future work will explore the scalability of generating DTs for more complex network topologies and traffic profiles, ensuring that the generated DT remains reliable across varying scenarios. Additionally, we seek to characterise the generalisability of the approach for dynamic environments, varying traffic types, and diverse network topologies.

\section{Conclusion}

In this paper, we have explored the feasibility of automatically generating digital twins for network testing, aligned to the ITU-T's autonomous network architecture \textit{experimentation subsystem}. We have presented how AutoML frameworks can be used to automatically generate DTs based on network data, addressing the human effort in the design and implementation of DTs. Initial experimental results have demonstrated the approach is feasible, even in the presence of noisy data, and provides high accuracy and very low execution times compared to equivalent simulators.
Future work will explore more complex network topologies and scenarios.

\section{Acknowledgment}
This work was supported by EPSRC funding EP/Z533221/1.


\bibliographystyle{ieeetr}
\bibliography{references}

\end{document}